\documentclass[12pt]{article}
\def\e{{\rm e}}
\def\ln{{\rm ln}} 
\title{\bf 
The Information Geometry\\ 
of the One-Dimensional \\
Potts Model
}
\author{ {\it B.P. Dolan}\\
Department of Mathematical Physics\\
National University of Ireland\\
Maynooth, Ireland\\
{}\\
{\it D.A. Johnston}\\ 
Dept. of Mathematics\\
Heriot-Watt University\\
Riccarton\\
Edinburgh, EH14 4AS, Scotland
\\
{\bf and}
\\
{\em R. Kenna}\\
School of Mathematical and Information Sciences\\
Coventry University\\
Coventry, CV1 5FB, England
}
\begin{document}
\maketitle
%-----------------------------------------------------------------------
                      {\Large
                      \begin{abstract}
%-----------------------------------------------------------------------
%
In various statistical-mechanical models the introduction of a metric 
onto the space of parameters
(e.g. the temperature variable, $\beta$, and the external field variable, $h$,
in the case of spin models)
gives an alternative perspective on the phase structure.
For the one-dimensional Ising model 
the scalar curvature, ${\cal R}$, of this metric can be calculated explicitly
in the thermodynamic limit and is found to be ${\cal R} = 1 + \cosh (h) / \sqrt{
\sinh^2 (h) + \exp ( - 4 \beta)}$. This is positive definite
and, 
for physical
fields and temperatures, diverges only
at the zero-temperature, 
zero-field
 ``critical point'' of the model.
In this note we calculate ${\cal R}$ for the one-dimensional $q$-state Potts model 
finding an expression of the form ${\cal R} = A(q,\beta,h) 
+ B (q,\beta,h)/\sqrt{\eta(q,\beta,h)}$,
where $\eta(q,\beta,h)$ is the Potts analogue of $\sinh^2 (h) + \exp ( - 4 \beta)$. This
is no longer positive definite, but once again it diverges only 
at the critical point
in the space of real parameters. We remark, however, 
that a naive analytic continuation to complex field
reveals a further divergence in the Ising and Potts 
curvatures at the Lee-Yang edge.  
%
%-----------------------------------------------------------------------
                        \end{abstract} }
%-----------------------------------------------------------------------
%
  \thispagestyle{empty}
%
%***********************************************************************
%
  \newpage
%
%-----------------------------------------------------------------------
                  \pagenumbering{arabic}
%-----------------------------------------------------------------------

%%%%%%%%%%%%%%%%%%%%%%%%%%%%%%%%%%%%%%%%%%%%%%%%%%%%%%%%%%%%%%%%%%

\section{Introduction}
%%%%%%%%%%%%%%%%%%%%%%%%%%%%%%%%%%%%%%%%%%%%%%%%%%%%%%%%%%%%%%%%%%

The notion of a distance between configurations in statistical-mechanical
and lattice-field models has been discussed by 
several authors \cite{Rupe,Jany,Brody,Brian,BrianA,Brianetc}, 
and the geometry that this endows on the manifold, ${\cal M}$, of the parameters
characterising the models explored. This distance is measured using the
Fisher-Rao metric \cite{Fish}, which is calculated
from the Fisher information matrix for the system of interest.
When we consider a spin model in field, ${\cal M}$
is a two-dimensional manifold parametrised by
$(\theta^{1},\theta^{2})=(\beta,h)$. In this case, the components of
the  Fisher-Rao metric take the particularly 
simple form 
\begin{equation}
G_{ij} = -\partial_{i}\partial_{j} f ,
\label{frmetric}
\end{equation}
where $f$ is the reduced free energy per site and
$\partial_{i} = \partial/\partial\theta^{i}$. 

It has been suggested that in such a geometrization of statistical
mechanics the scalar curvature, ${\cal R}$, of ${\cal M}$ plays a central role
\cite{Rupe, Jany, Brody}. 
For a spin model in field,
with the metric given in equ.(\ref{frmetric}),
${\cal R}$ may be calculated succinctly as
\begin{equation} 
{\cal R}\ =\frac{1}{2 G^{2}} 
\left| \begin{array}{lll} 
\partial^{2}_{\beta} f & \partial_{\beta}\partial_{h}f & 
\partial_{h}^{2}f \\ 
\partial^{3}_{\beta}f & \partial_{\beta}^{2}\partial_{h}f & 
\partial_{\beta}\partial_{h}^{2}f \\ 
\partial^{2}_{\beta}\partial_{h}f & 
\partial_{\beta}\partial_{h}^{2}f & \partial_{h}^{3}f  
\end{array} \right| \ , 
\label{equcurv} 
\end{equation}
where $G={\rm det}(G_{ij})$. Since the only scale present near criticality for
a model displaying a higher-order transition is the correlation length,
$\xi$,
it has been hypothesised on dimensional grounds that ${\cal R} \sim \xi^d$,
where $d$ is the dimensionality of the system \cite{Rupe, Jany, Brody}.
If we assume that
hyperscaling holds, $\nu d = 2 - \alpha$,  this leads to
\begin{equation}
{\cal R} \sim |\xi|^{ ( 2 - \alpha)/\nu}.
\label{equscal}
\end{equation}

To test the behaviour of ${\cal R}$, one requires models which are
solvable in field in order to obtain analytic expressions, and these are rather
thin on the ground. Indeed, ${\cal R}$ has been calculated for the mean-field
and
Bethe-lattice
Ising models \cite{Brian} and the above scaling behaviour verified. It has also
been calculated for the one-dimensional Ising model \cite{Jany} where it takes
the remarkably simple form,
\begin{equation}
{\cal R}_{\rm{Ising}}=1 + {\cosh h \over \sqrt{\sinh^{2}h+e^{-4\beta}}}.
\label{RIsing}
\end{equation}
In this case ${\cal R}$ is positive definite
and diverges only at the zero-temperature, 
zero-field
``critical point'' 
of the model. The correlation length
is given by
\begin{equation}
 \xi^{-1} = -\ln{\left(\tanh (\beta)\right)},
\end{equation}
so that $\xi \sim  \exp(2\beta)$ near criticality, 
and (\ref{equscal}) holds there with $\alpha=1, \nu=1$ as 
expected\footnote{The Bethe lattice model
also satisfies the postulated scaling, although there are some subtleties
coming from the exponent $\alpha$ being zero \cite{Brian}.}. 
A second noteworthy feature of the 
Bethe-lattice and 1D Ising models
concerns the metric associated with (\ref{frmetric}).
This metric is diagonalized using the corresponding 
renormalization group invariant.

Given the shortage of explicitly calculable examples, any further additions to the 
list would be very worthwhile in order to see which features in the models 
are generic and which are particular to the models concerned. In this paper we
discuss another example, the 1D $q$-state Potts model, where an expression
for ${\cal R}$ may be obtained in a very similar manner to the Ising model
(which is the $q=2$ case of the Potts model).
We compare the properties of the 
curvature,
${\cal R}$, as well as the metric in the Potts model and the
Ising model, highlighting both their structural similarities and differences
in detail. 

%%%%%%%%%%%%%%%%%%%%%%%%%%%%%%%%%%%%%%%%%%%%%%%%%%%%%%%%%%%%%%%%%%

\section{The 1D Potts Model}
%%%%%%%%%%%%%%%%%%%%%%%%%%%%%%%%%%%%%%%%%%%%%%%%%%%%%%%%%%%%%%%%%%

The partition function for the 1D $q$-state Potts 
model is given by
\begin{equation}
Z_N (y, z) =\sum_{\{\sigma\}} \exp \left[ {\beta \sum_{j=1}^N \left( \delta( \sigma_j,  \sigma_{j+1})  - {1 \over q} \right) +  h
\sum_{j=1}^N \left( \delta ( \sigma_j, 1) - {1 \over q} \right)} \right]
,
\label{ZPotts}
\end{equation}
where the spins, $\sigma_j\in \{1,\dots, q\}$, are defined on the
sites, $j \in \{1,\dots N\}$, of the lattice
and where
we have defined $y = \exp ( \beta )$ and $z = \exp ( h )$ for later calculational convenience.
The model may be solved by transfer matrix methods \cite{Glue}, just as the 1D Ising model. 
For general $q$ the full transfer matrix $T ( y, z)$ 
may be written as $q-2$ diagonal elements, 
$( y - 1) (  y z )^{-1/q}$, and
a $2 \times 2$ factor $t ( y, z)$:
\begin{equation}
t ( y, z) = {1 \over ( y z )^{1/q}  }\pmatrix{
 y z & z^{1/2} \sqrt{ q - 1} \cr
 z^{1/2}\sqrt{q - 1}  & y + q - 2 \cr}.
\label{eP}
\end{equation}
The partition function is $Z_N (y, z) = Tr T (y, z)^N$ 
and the eigenvalues of 
$T(y,z)$ are $\lambda_0,\lambda_1,\dots,\lambda_{q-1}$, where
\begin{equation}
\left.
\begin{array}{ll}
\lambda_{0}\\
\lambda_{1}
\end{array}
\right\}
 = {1 \over 2} \left[{  y ( 1 +z ) + q - 2    
\pm 
\sqrt{ (y ( 1 - z) + q - 2)^2 + ( q- 1) 4 z } \, }\right] \, ( y z ) ^{ - {1 \over q}}
\label{eigen}
\end{equation}
and $\lambda_2, \dots \lambda_{q-1} = ( y - 1 ) (y z ) ^{ - {1 / q}}$.
The reduced free energy per site in the thermodynamic limit,
$N\rightarrow \infty$, is thus given by $f = -\ln \lambda_{0}$.

It is straightforward to use this expression for the  free energy in equ.(\ref{equcurv})
to obtain the curvature, ${\cal R}$. In the current notation we re-derive the expression for the Ising model \footnote{There is a factor
of two difference in the definitions of $\beta, h$  between
the Ising and Potts notations coming from the
different spin definitions.}
as
\begin{equation}
{\cal R}_{\rm{Ising}}=1 + {y ( 1 + z) \over \sqrt{y^2 - 2 y^2 z + y^2 z^2 +4 z}}.
\end{equation}
The expression for general $q$ is similar in form to this, and is
\begin{equation}
{\cal R}_{\rm{Potts}} = A (q, y, z)   + {B(q, y , z) \over  \sqrt{\eta (q, y, z)}},
\end{equation}
where the coefficients may be further broken down as
$A (q, y, z) = \alpha(q, y, z) / \gamma(q, y , z)^2$ and $B(q, y, z) = \beta (q, y , z) / \gamma(q, y , z)^2$
and are smooth functions of $y$ and $z$ and do not diverge for finite (physical) temperature
or field. Furthermore
\begin{equation}
\eta (q, y, z) =  [ y ( 1 - z) + q - 2 ]^2 + ( q- 1) 4 z .
\label{eta}
\end{equation}
The expressions for $\alpha(q, y, z),\, \beta (q, y , z)$ and $\gamma(q, y , z)$ are very lengthy for general $q$ (although still easily obtained)
so, for the sake of compactness,
we write down only those for $q=3$, which already display
the generic features. We have
%%%%%%%%%%%%%%%%%%%%%%%%%%%%%%%Big equs start here%%%%%%%%%%%%%%%%%%%%%%%%%%%%%%%%%
\begin{eqnarray}
\lefteqn{
\alpha( 3 , y , z) 
=  - \frac{1}{4}  
\left[
 - 6 
 - 12\,z 
+ 116\,z^{2}
+  80\,z^{3} 
-  16\,z^{4} 
\quad \quad \quad
\right.
}
\nonumber \\  
& & + y
      ( -  44 
        -  68\,z 
        + 400\,z^{2} 
        + 552\,z^{3} 
        - 192\,z^{4} 
      )
\nonumber \\ 
& & +y^2
      ( - 113 
        -  64\,z 
        + 138\,z^{2} 
        + 476\,z^{3}
        -1004\,z^{4}
      )
\nonumber \\ 
& & + y^3
      ( - 178
        +  30\,z 
        + 384\,z^{2} 
        -1884\,z^{3} 
        -2896\,z^{4} 
        +   8\,z^{5} 
      )
\nonumber \\ 
& & + y^4
      ( - 188
        + 458\,z 
        - 222\,z^{2} 
        - 276\,z^{3} 
        -3400\,z^{4} 
        +  64\,z^{5} 
      )
\nonumber \\ 
& & +y^5
      ( - 106
        + 592\,z 
        - 474\,z^{2} 
        -  68\,z^{3}
        +1232\,z^{4} 
        + 120\,z^{5} 
      )
\nonumber \\ 
& & + y^6
      (
        -  17
        + 294\,z 
        - 368\,z^{2} 
        + 250\,z^{3} 
        - 127\,z^{4}
        -  32\,z^{5}
      )
\nonumber \\ 
& & + y^7
\left.
      (
            4
        +  66\,z 
        - 136\,z^{2} 
        +  60\,z^{3}
        +   2\,z^{5} 
        +   4\,z^{4}
      )
\right],
\quad \quad \quad \quad\quad \quad\quad \quad\quad 
\end{eqnarray}
%%\\
%%{\mbox{and}}
%%\nonumber \\
and
\begin{eqnarray}
\lefteqn{
\beta ( 3, y, z) 
= 
- \frac{1}{4} 
\left[
-   30 
+   12\,z 
+  468\,z^{2} 
+  768\,z^{3} 
+  240\,z^{4} 
\right.
}
\nonumber \\
& & 
+ y
(
-  82
-  82\,z 
+ 664\,z^{2} 
+3604\,z^{3} 
+1744\,z^{4} 
-  16\,z^{5} 
)
\nonumber \\
& & 
+y^2
(
- 171
+  12\,z 
+1366\,z^{2}
+2292\,z^{3} 
+1764\,z^{4} 
- 160\,z^{5} 
)
\nonumber \\ 
& & 
+y^{3}
(
- 261
+ 213\,z 
+1710\,z^{2} 
+ 846\,z^{3} 
-7584\,z^{4} 
- 756\,z^{5} 
)
\nonumber \\ 
& & 
+y^4
(
- 278
+ 386\,z 
+ 400\,z^{2} 
-3804\,z^{3}  
-8804\,z^{4} 
-2488\,z^{5}
+   8\,z^{6} 
)
\nonumber \\ 
& & 
+y^5
(
- 254
+ 570\,z 
-1094\,z^{2}
-5654\,z^{3} 
-1584\,z^{4}
-3712\,z^{5}
+  64\,z^{6} 
)
 \nonumber \\ 
& & 
+y^6
(
- 165
+ 504\,z 
-1494\,z^{2} 
-1008\,z^{3} 
+ 891\,z^{4}
+1152\,z^{5} 
+ 120\,z^{6}
)
\nonumber \\
& & 
+ y^7
(
-  51
+ 263\,z
- 866\,z^{2}
+ 986\,z^{3} 
- 203\,z^{4}
-  97\,z^{5}
-  32\,z^{6} 
)
\nonumber \\ 
& & 
+y^8
\left.
(
-   4
+  66\,z 
- 182\,z^{2} 
+ 188\,z^{3} 
-  72\,z^{4}
+   2\,z^{5}
+   2\,z^{6} 
)
\right],
\end{eqnarray} 
for the numerators, and
\begin{eqnarray}
\gamma(3, y, z) = 
&-& 3 - 4\,z - 2\,z^{2} - 4\,y - 16\,y\,z - 16\,y\,z^{2}  \nonumber \\
&-&   7\,y^{2
} + 2\,y^{2}\,z - 31\,y^{2}\,z^{2} - 4\,y^{3} + 4\,y^{3}\,z^{2}
\, ,
\end{eqnarray}
and
\begin{eqnarray}
\eta (3, y, z) =  
1 + 8\,z + 2\,y - 2\,y\,z
+ y^{2} - 2\,y^{2}\,z + y^{2}\,z^{2},
\end{eqnarray}
for the terms in the denominators.
%%%%%%%%%%%%%%%%%%%%%%%%%%%%%%%Big equs end here%%%%%%%%%%%%%%%%%%%%%%%%%%%%%%%%%

In zero-field ($z=1$) the 
expression for ${\cal R}$  is much more compact
and is  written for general $q$ as
\begin{equation}
{\cal R}  = { (y + q -1 ) ( 4 y^2 + ( q - 2 ) y - (q - 2) ( q  - 1)) \over (q - 1) ( 2 y + q  - 2 )^2 }.
\label{Rz1}
\end{equation}
We see that as $y$ ranges from $1$ to $\infty$, ${\cal R}$ ranges
from $(4 - q) / ( q -1)$ to $\infty$. In particular, the sign of the
$y=1$ ($\beta=0$) limit of ${\cal R}$ changes at $q=4$, although
the general morphology of ${\cal R}$ 
as  a function of $y$ and $z$ remains the same for all $q>2$
as we see below.

The correlation length for the one-dimensional Potts model
is defined in a similar manner to that of the Ising model,
\begin{equation}
 \xi^{-1} = -\ln{\left(  {\lambda_1 \over \lambda_0} \right)},
\end{equation}
so $\xi \sim y$ for $z=1, y \rightarrow \infty$. 
We thus retrieve the expected scaling  of
${\cal R}$ 
for the one-dimensional
Potts model
from equ.(\ref{equscal}), 
namely ${\cal R} \sim y$ as $y \rightarrow \infty$. 
The exponents, as for the Ising model,
are $\alpha=1, \nu=1$.

The general features of ${\cal R}$ at non-zero temperature and field
are perhaps easiest seen in a 
contour plot as a function of $y$ and $z$. In Fig.1 we show the Ising ($q=2$) case
which has certain non-generic features. The  $\pm h$ symmetry
of the Ising model is manifest as a $z \rightarrow 1/z$ symmetry
in the plot of ${\cal R}$. In addition, one can see that ${\cal R}$
is positive for all $y$ and $z$. The maximum of ${\cal R}$ 
for a given $y$ value lies
along the zero field line at $z=1$.

In Fig.2, ${\cal R}$ is plotted for the $3$-state 
Potts model, using the expressions given above for $A(3,y,z)$
and $B(3,y,z)$. We see that there is no longer a $z \rightarrow 1/z$ symmetry
and that ${\cal R}$  is {\it not}  positive definite. This behaviour is
typical of the $q>3$ case, as can also be seen from the $q=10$ results plotted in Fig.3.
It is also clear from Fig.3. that the $z=1, y=1$ limit is negative
unlike the $3$-state model,
as indicated by equ.(\ref{Rz1}). When $q > 2$ the maximum
of ${\cal R}$ no longer lies on the zero-field line though
its locus is still simply determined, as we discuss
in the next section.

%%%%%%%%%%%%%%%%%%%%%%%%%%%%%%%%%%%%%%%%%%%%%%%%%%%%%%%%%%%%%%%%%%

\section{Remarks on Co-ordinates and Geodesics}
%%%%%%%%%%%%%%%%%%%%%%%%%%%%%%%%%%%%%%%%%%%%%%%%%%%%%%%%%%%%%%%%%%

We have concentrated so far on the curvature ${\cal R}$, which is
of course a geometric invariant and independent of the particular co-ordinate
scheme we use in the calculation. The properties of the particular
metrics and line elements used in the calculation are also of some
interest \cite{BrianA}. The metric for the 1D Ising model
calculated from equ.(\ref{frmetric}) gives the line element
\begin{eqnarray}
ds^2 &=&
  {\e^{-4 \beta}\over\bigl(\sinh^2h+\e^{-4 \beta}\bigr)^{3/2}}
      \times  \left[
         {4\e^{-4 \beta}\cosh h + 8\sinh^2 h
             \bigl(\cosh h+\sqrt{\sinh^2h+\e^{-4 \beta}}\bigr)
         \over{\bigl(\cosh h+\sqrt{\sinh^2h+\e^{-4 \beta}}
               \bigr)^2}}\ d\beta^2\right. \nonumber \\
        &+&  \left. 
         4\sinh h \ d\beta dh + \cosh h \ dh^2\right].
\end{eqnarray}
Although this is clearly non-diagonal, 
the choice of $\beta$ and $h$ as 
co-ordinates has the advantage of giving the simplified expression,
 equ.(\ref{equcurv}), for ${\cal R}$ since they appear as linear
multipliers in the Hamiltonian.

The metric can be diagonalized by inspection in this case, and this 
preferred choice of co-ordinates turns out to be rather interesting from the
physical point of view. If one considers a decimation type renormalization
transformation on a 1D Ising chain (or ring)
where every other vertex is decimated and the length scale
suitably adjusted, then one can exactly
determine the relations between the renormalized parameters
$\beta^\prime, h^\prime$ and the originals $\beta, h$ from
\begin{equation}
Z_{N / 2} ( \beta^\prime, h^\prime ) = A^N Z_N ( \beta, h )
\end{equation}
where $Z_{N / 2}$ is the decimated partition function and
$Z_N ( \beta, h )$ is the original one. In addition to the parameter
transformation there is also an unimportant overall scaling 
factor, $A^N$.
This transformation has an invariant
$\rho=\e^{2 \beta}\sinh h$, which is in effect encoding the
invariance of the magnetization under the transformation 
since 
\begin{equation}
M = { \rho \over \sqrt{1 + \rho^2}}.
\end{equation}
If the co-ordinates are transformed from $\beta, h$ to $\beta, \rho$
then, using
\begin{equation}
d\rho = 2 \e^{2 \beta}\sinh h d \beta + \e^{2 \beta}\cosh h dh,
\end{equation}
one finds  the diagonalized expression
\begin{eqnarray} 
ds^2={1\over\sqrt{(1+\rho^2)(\e^{4 \beta} +\rho^2)}}
       \left[
   {4\;\e^{4 \beta} d\beta^2\over
       \bigl(\sqrt{1+\rho^2}+\sqrt{\e^{4 \beta}+\rho^2}\bigr)^2}
         + {d\rho^2\over (1+\rho^2)}\right].
\end{eqnarray}

The use of the renormalization group invariant, or alternatively
the magnetization, also diagonalizes the metric for the Ising
model on a Bethe lattice. It is therefore natural to ask whether
this feature persists in the Potts models considered here.
Although the Potts metric and line elements are rather
more complicated than those for the Ising model, the same general
structure is
once again apparent
\begin{eqnarray}
ds^2 &=& {( q - 1) \over \eta ( q, y, z)^{(3/2)} }
\times  \left[ { C ( q, y, z) + D (q, y, z) \sqrt{\eta ( q, y, z)}  \over
\tilde 
\lambda( q, y, z)^2} d \beta^2 +  4 { [ y z (  z -1 )  ]  } d \beta d h \right.
\nonumber \\
&+& \left. { z [ y ( 1 + z) + q -2 ]  } dh^2 \right]
\label{betahmetric}
\end{eqnarray}
where $\eta ( q, y, z)$ is  defined in equ.(\ref{eta}) and
\begin{equation}
\tilde \lambda( q, y, z) = {  y ( 1 +z ) + q - 2
+
\sqrt{ (y ( 1 - z) + q - 2)^2 + ( q- 1) 4 z } \, }
\end{equation}
is proportional to $\lambda_0$, the dominant eigenvalue.
The functions $C ( q, y, z)$ and $D (q, y, z)$ are both
easily calculable, but rather long, and are not reproduced here.

If we carry out a decimation on the 1D Potts model
\begin{equation}
Z_{N / 2} ( y^\prime, z^\prime ) = A^N Z_N ( y, z ),
\end{equation}
then the invariant  is given by
\begin{equation}
\tilde \rho = { [ y ( 1 - z) + q - 2 ] \over \sqrt{z} },
\label{Pottsinv}
\end{equation}
which is again related to the magnetization.
Following the Ising procedure, we can change variables
$\beta, h$ to $\beta, \tilde \rho$ using
\begin{equation}
d \tilde \rho =  { y ( 1 - z) \over \sqrt{z} } d \beta - {1 \over 2} { [ y(1+z) + q-2 ] \over \sqrt{z} } 
dh
\end{equation} 
and find that the metric is, indeed, diagonalized. 

In the Ising model it can be shown that the line $\rho=0$ (i.e. $h=0$
or $z=1$) is a
geodesic of the metric (\ref{frmetric}). In Fig.1 this is the line running along the ridge
of ${\cal R}$. 
For $q \ne 2$ Potts models the local maximum in ${\cal R}$ no longer lies
at $z=1$, as is clear
from Fig.2 and Fig.3, however an analysis of the geodesic equations
\begin{eqnarray}
{d V^\beta\over ds} + \Gamma^\beta_{\beta\beta}V^\beta V^\beta
+ 2\Gamma^\beta_{\beta h}V^\beta V^h + \Gamma^\beta_{h h }V^h V^h
& = & \lambda(s) V^\beta \nonumber \\
{d V^h\over ds} + \Gamma^h_{\beta\beta}V^\beta V^\beta
+ 2\Gamma^h_{\beta h}V^\beta V^h + \Gamma^h_{h h }V^h V^h 
& = & \lambda(s) V^h
\label{geodesic}
\end{eqnarray}
where $s$ parameterizes the flow
lines, $V^\beta = {d \beta / ds}$, $V^h = {d h / ds}$,
the $\Gamma$ are the various Christoffel symbols for the metric
of equ.(\ref{frmetric}),
and $\lambda(s)$ allows for non-affine parameters,
shows that the line $z=1$ is still a geodesic. 
To prove this consider a vector field with a flow line along $z=1$ ($h=0$).
This flow line has $V^h=0$ so equations (\ref{geodesic})
become
\begin{eqnarray}
{d V^\beta\over ds} + 
\left(\Gamma^\beta_{\beta\beta}\right)\vrule height14pt depth6pt\raise -4pt \hbox{$_{h=0}$}V^\beta V^\beta
&=&\lambda(s) V^\beta \\
\left(\Gamma^h_{\beta\beta}\right)\vrule height14pt depth6pt\raise -4pt \hbox{$_{h=0}$}V^\beta V^\beta
& = & 0.
\end{eqnarray}
The first of these equations is a second order 
ordinary differential equation for $\beta(s)$ which always
has a solution.  The second requires 
$\Bigl(\Gamma^h_{\beta\beta}\Bigr)\vrule height12pt depth4pt\,\raise -2pt \hbox{$_{h=0}$}=0$
and it can easily be shown, using the metric (\ref{betahmetric}), that this is indeed the case.
Hence the line $z=1$ is a geodesic of the metric (\ref{betahmetric}) for any value of $q$.

A different change of variable in the Potts models
brings one back to something
very similar to the Ising picture. If we transform $z$ to
\begin{equation}
w = {y z \over y + q - 2 }  
\end{equation}
and leave $y$ untouched, we find that the local maximum in ${\cal R}$
lies on the line $w=1$. We show ${\cal R}$ plotted against $w$
in Fig.4 for the 3-state Potts model for comparison with Fig.2.
The picture is similar for  higher $q$. This transformation does
not greatly simplify $\alpha(q,y,z)$, $\beta(q,y,z)$ or $\gamma(q,y,z)$
and we do not reproduce the expressions here.

In summary, although the choice of $\beta, h$ as parameters is natural
in any calculation of ${\cal R}$, the metric is diagonalized for
both the 1D Ising and Potts models when the renormalization group
invariant, or equivalently the magnetization, is used as a 
co-ordinate instead 
of $h$. The line $z=1$ is a geodesic of the metric (\ref{frmetric}) for any $q$.

%%%%%%%%%%%%%%%%%%%%%%%%%%%%%%%%%%%%%%%%%%%%%%%%%%%%%%%%%%%%%%%%%%

\section{A Lee-Yang Divergence}
%%%%%%%%%%%%%%%%%%%%%%%%%%%%%%%%%%%%%%%%%%%%%%%%%%%%%%%%%%%%%%%%%%

Some years ago Lee and Yang \cite{YL} 
addressed the question of how the singularities
associated with field-driven phase transitions in Ising-like spin models 
on lattices
arose in the thermodynamic limit. 
This was later extended by various authors 
to other models and to temperature-driven transitions \cite{others,fish}.
Lee and Yang
observed that the zeroes of the partition function
for a spin model in a complex external field
on a finite lattice would give rise
to singularities in the free energy. 
In the thermodynamic limit 
these complex zeroes
move in to pinch the real axis, signalling the the onset of a
physical
phase transition.
Typically, the loci of zeroes are lines in the complex field
or temperature plane
and when the
endpoints of such lines
occur at non-physical (i.e. complex) external field values they can
be considered as ordinary critical points with an associated edge critical
exponent, usually dubbed the Lee-Yang edge exponent \cite{others}. 

The Lee-Yang zeroes for the one-dimensional
Potts model on a periodic chain with $N$ sites
are given by the solutions \cite{Glue, Brian2} of
\begin{equation}
Z_N = (\lambda_1)^N +(\lambda_0)^N = 0 \qquad\Leftrightarrow \qquad
\lambda_1 = \exp( {in\pi\over N} )
\lambda_0    
\end{equation}
where $\lambda_{0,1}$  are the eigenvalues given in 
equ.~(\ref{eigen}) and $n$ is odd. In the thermodynamic limit
the locus of zeroes is determined by $| \lambda_0 | = | \lambda_1 |$
or
\begin{equation}
\eta(q, y, z) = [y ( 1 - z) + q - 2]^2 + ( q- 1) 4 z  = 0
\end{equation}
which can be satisfied for complex (in
the $q=2$ Ising case, purely imaginary) values of  $h$.

Unlike the Ising model case,  
the Lee-Yang zeroes for the 1D $q \ne 2$ Potts models do not lie
on the unit circle in the complex $z$ plane. However, precisely
the change of co-ordinates we used in discussing the Potts
metrics in the previous section, 
$w =  y z / ( y + q - 2 )$, places them on the unit circle in 
the complex $w$
plane. If we associate a ``field'' with $w$ via $w= exp( \tilde h)$,
then purely imaginary values of $\tilde h$ give the zeroes, as in 
the Ising case\footnote{This is not, of course, the field $h$
appearing in the Hamiltonian.}. 

From these considerations, 
it is clear that
${\cal R}$ will also diverge as
the locus of zeroes is approached 
for both Ising and Potts
models if we allow ourselves the liberty of an analytic
continuation of the field to complex $h$ values once 
${\cal R}$
has been calculated,
since ${\cal R} = A + B / \sqrt{\eta}$ 
and $A, B$ are  finite as $\eta \rightarrow 0$.
The presence of the square root means that the  
divergence is characterised by an exponent $\sigma=-1/2$ which is 
the Lee-Yang edge exponent for the one-dimensional
Potts (and Ising) model \cite{others}. 

The status of these
observations are a little unclear to us, since the calculation
of ${\cal R}$ has assumed a real metric geometry throughout and such
an arbitrary continuation in the final
expression might be rather dangerous. 
One is on slightly firmer ground with the Ising model,
since in that case the required continuation is to purely
imaginary fields which correspond simply
to a change in the signature of the metric in $\beta, \rho$
co-ordinates. 
The use of the $w$ co-ordinate in the Potts case does,
however, suggest
a similar interpretation since there the zeroes occur
 for imaginary values of $\tilde h$.
With these caveats,
it is nonetheless interesting that the Lee-Yang edge transition 
is still visible as a divergence in ${\cal R}$. 

%%%%%%%%%%%%%%%%%%%%%%%%%%%%%%%%%%%%%%%%%%%%%%%%%%%%%%%%%%%%%%%%%%

\section{Conclusions}
%%%%%%%%%%%%%%%%%%%%%%%%%%%%%%%%%%%%%%%%%%%%%%%%%%%%%%%%%%%%%%%%%%

We have seen that the scalar curvature, ${\cal R}$, of the
Fisher-Rao metric may be obtained for the 1D $q$-state Potts model
in a very similar fashion to the 1D Ising model
since the free energy can be calculated in field in both cases
using transfer matrix methods.

Although ${\cal R}$ for the $q$-state Potts model cannot be expressed
as succinctly  as for the Ising model it still has the same general 
form, ${\cal R} = A + B / \sqrt{\eta}$, and diverges only at
the zero temperature critical point of the Potts model
for physical temperature and field values, i.e.
$y = 1 \ldots \infty$ and $z=0 \ldots \infty$. We 
wrote down  ${\cal R}$ explicitly here for $q=3$, but noted it was a simple
matter to calculate it for arbitrary $q$.
We observed that 
there were some features
of the Ising model ${\cal R}$ which did 
not persist for general $q$, as was clear from contour plots. In particular,
${\cal R}$ was no longer  positive definite
and the $z \rightarrow 1/z$ symmetry of the Ising model was
no longer present.

The choice of co-ordinate scheme used in the calculation of ${\cal R}$
was also discussed. We noted that although $\beta, h$ were a natural
choice from the point of view  of simplifying this calculation,
the use of the renormalization group invariant in place of $h$ diagonalized
the metric for both the Ising and Potts models. A different choice
of coordinates, involving a rescaling of $z$ was also employed to
identify the local maximum in ${\cal R}$.
For any value of $q$ the line $z=1$ ($h=0$) is a geodesic of the
Fisher-Rao metric.
Finally, we observed that if
complex field values are permitted
there is a divergence in ${\cal R}$ at the Lee-Yang edge with
an exponent $\sigma=-1/2$.

We have confined our discussion here to physical values of 
$q$, i.e. $2, 3, \dots$, in the Potts models. Since $q$ appears as
a parameter in the transfer matrix solution there is in principle no barrier
to extending it to general non-integer $q$ and also to $q<2$. 
The $q \rightarrow 1$ limit of the Potts model is related to percolation,
for example, so the behaviour of ${\cal R}$ might be of interest
in this limit too. It would also be interesting to calculate  ${\cal R}$ 
for models with genuine transitions at finite $\beta$, an exercise that
has so far been confined to mean-field or mean-field-like spin models.
Possibilities which suggest themselves in this context are
the Ising model on planar random graphs \cite{BK} and the spherical model.
  
%%%%%%%%%%%%%%%%%%%%%%%%%%%%%%%%%%%%%%%%%%%%%%%%%%%%%%%%%%%%%%%%%%

\section{Acknowledgements}
%%%%%%%%%%%%%%%%%%%%%%%%%%%%%%%%%%%%%%%%%%%%%%%%%%%%%%%%%%%%%%%%%%

D.J. was partially supported by
EC IHP network
``Discrete Random Geometries: From Solid State Physics to Quantum Gravity''
{\it HPRN-CT-1999-000161}. D.J. and R.K. were also partially supported
by an Enterprise Ireland/British Council Research Visits Scheme grant.

%%%%%%%%%%%%%%%%%%%%%%%%%%%%%%%%%%%%%%%%%%%%%%%%%%%%%%%%%%%%%%%%%%

\bigskip
%

%------------------------------------ 

\clearpage \newpage
\begin{figure}[t]
\vskip 15.0truecm
\includegraphics{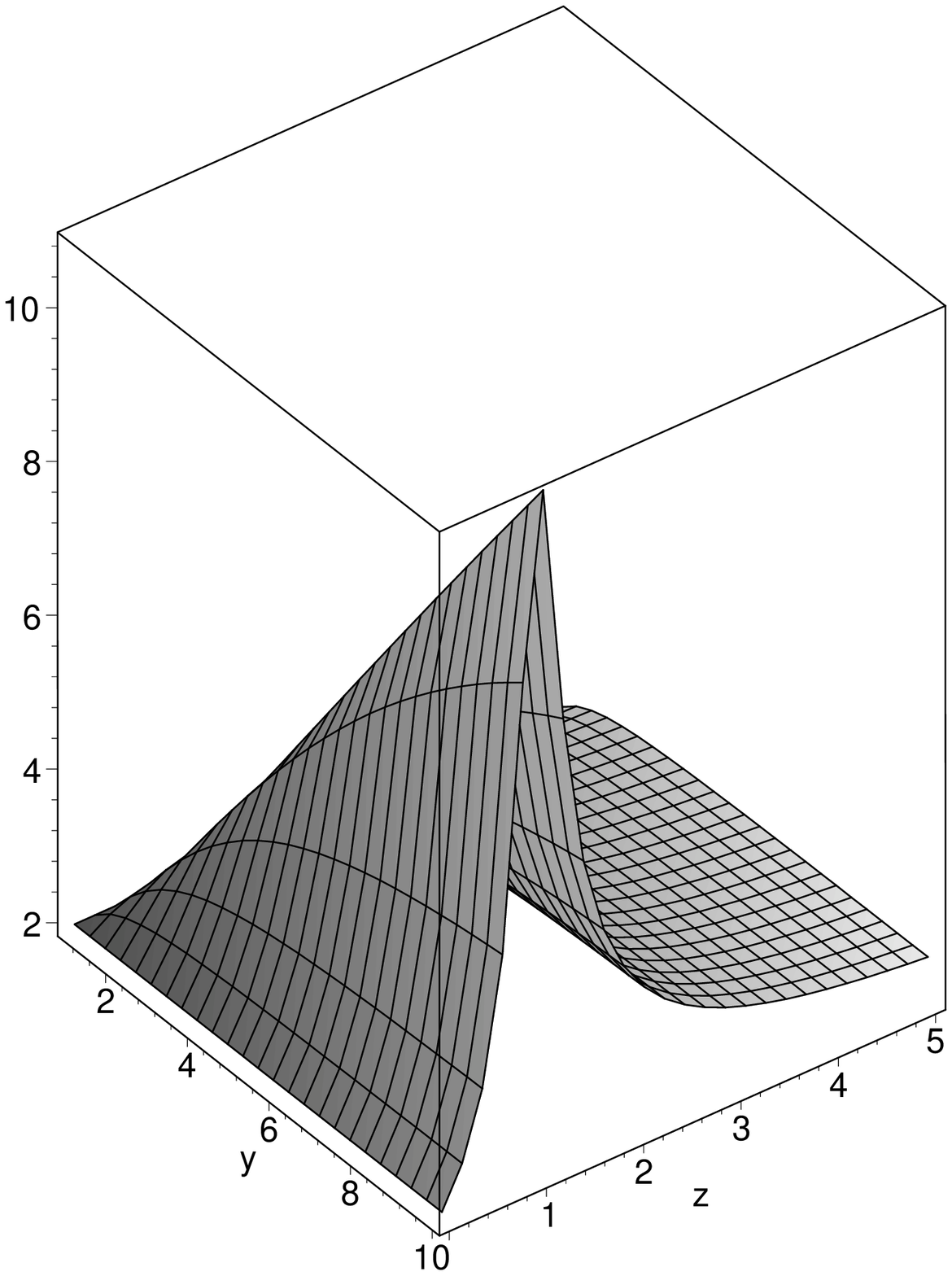}
\caption[]{\label{fig1} A plot of ${\cal R}$ for the Ising model
for $y=1 \ldots 10$, $z=0 \ldots 5$. The positivity of ${\cal R}$ and
the expected $z \rightarrow 1/z$ symmetry are both apparent.
}
\end{figure}

%------------------------------------

\clearpage \newpage
\begin{figure}[t]
\vskip 15.0truecm
\includegraphics{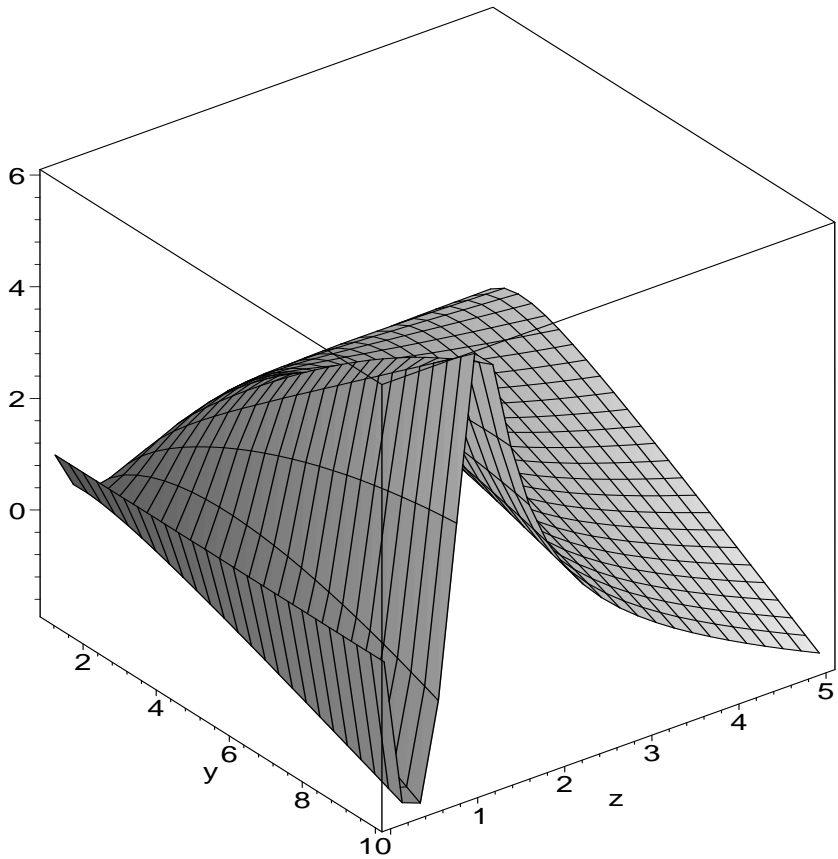}
\caption[]{\label{fig2} A plot of ${\cal R}$ for the $3$-state Potts model
for $y=1 \ldots 10$, $z=0 \ldots 5$. ${\cal R}$ is no longer
positive definite for physical values of $y,z$ and
there is no $z \rightarrow 1/z$ symmetry. In addition the maximum of
 ${\cal R}$ does not lie at $z=1$, though this is perhaps clearer for
higher $q$.
}
\end{figure}

%------------------------------------

\clearpage \newpage
\begin{figure}[t]
\vskip 15.0truecm
\includegraphics{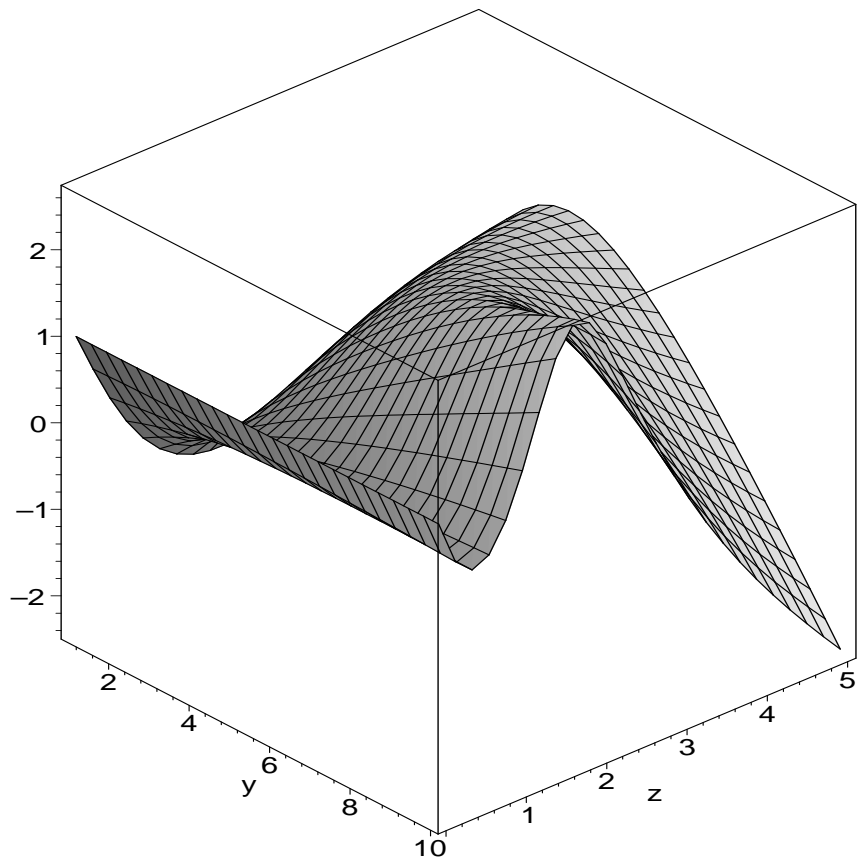}
\caption[]{\label{fig3} A plot of ${\cal R}$ for the $10$-state Potts model
for $y=1 \ldots 10$, $z=0 \ldots 5$. In this case it is quite clear that
the maximum of
 ${\cal R}$ does not lie at $z=1$.
}

\end{figure}

%------------------------------------
\clearpage \newpage
\begin{figure}[t]
\vskip 15.0truecm
\includegraphics{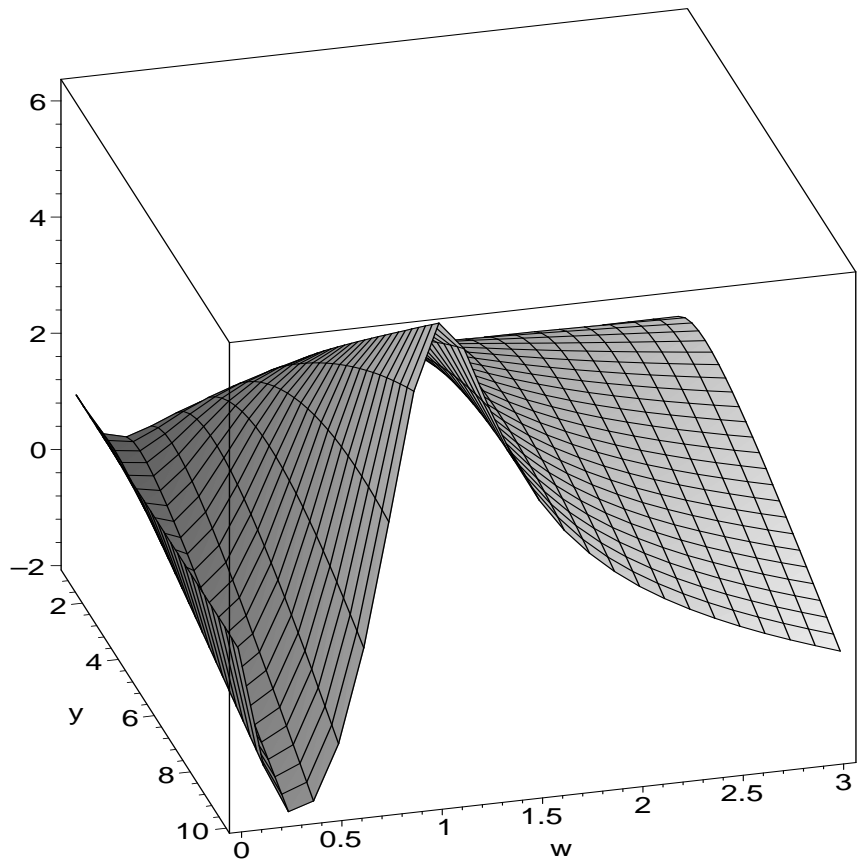}
\caption[]{\label{fig4} A plot of ${\cal R}$ for the $3$-state Potts model
for $y=1 \ldots 10$, $w=0 \ldots 3$. Note that the maximum of ${\cal R}$
 lies at $w=1$
}
 
\end{figure}

%------------------------------------
\end{document}